\documentclass[twocolumn]{aastex62}
\usepackage{graphicx}
\usepackage{color}
\usepackage{amsmath}
\begin{document}

\title{Search for Eccentric Binary Black Hole Mergers with Advanced LIGO and Advanced Virgo \\ during their First and Second Observing Runs}
%\author{LIGO Scientific Collaboration, Virgo Collaboration and F. Salemi}
\author{B.~P.~Abbott}
\affiliation{LIGO, California Institute of Technology, Pasadena, CA 91125, USA}
\author{R.~Abbott}
\affiliation{LIGO, California Institute of Technology, Pasadena, CA 91125, USA}
\author{T.~D.~Abbott}
\affiliation{Louisiana State University, Baton Rouge, LA 70803, USA}
\author{S.~Abraham}
\affiliation{Inter-University Centre for Astronomy and Astrophysics, Pune 411007, India}
\author{F.~Acernese}
\affiliation{Dipartimento di Farmacia, Universit\`a di Salerno, I-84084 Fisciano, Salerno, Italy}
\affiliation{INFN, Sezione di Napoli, Complesso Universitario di Monte S.Angelo, I-80126 Napoli, Italy}
\author{K.~Ackley}
\affiliation{OzGrav, School of Physics \& Astronomy, Monash University, Clayton 3800, Victoria, Australia}
\author{C.~Adams}
\affiliation{LIGO Livingston Observatory, Livingston, LA 70754, USA}
\author{R.~X.~Adhikari}
\affiliation{LIGO, California Institute of Technology, Pasadena, CA 91125, USA}
\author{V.~B.~Adya}
\affiliation{OzGrav, Australian National University, Canberra, Australian Capital Territory 0200, Australia}
\author{C.~Affeldt}
\affiliation{Max Planck Institute for Gravitational Physics (Albert Einstein Institute), D-30167 Hannover, Germany}
\affiliation{Leibniz Universit\"at Hannover, D-30167 Hannover, Germany}
% ...

\collaboration{(The LIGO Scientific Collaboration and the Virgo Collaboration)}

\author{F.~Salemi}
\affiliation{Max Planck Institute for Gravitational Physics (Albert Einstein Institute), D-30167 Hannover, Germany}

\begin{abstract}
When formed through dynamical interactions, stellar-mass binary black holes may retain eccentric orbits ($e>0.1$ at 10\,Hz) detectable by ground-based gravitational-wave detectors. Eccentricity can therefore be used to differentiate dynamically-formed binaries from isolated binary black hole mergers. Current template-based gravitational-wave searches do not use waveform models associated to eccentric orbits, rendering the search less efficient to eccentric binary systems. Here we present results of a search for binary black hole mergers that inspiral in eccentric orbits using data from the first and second observing runs (O1 and O2) of Advanced LIGO and Advanced Virgo. {We carried out the search with the coherent WaveBurst algorithm, which uses minimal assumptions on the signal morphology and does not rely on binary waveform templates.} We show that it is sensitive to binary mergers with a detection range that is weakly dependent on eccentricity for all bound systems. Our search did not identify any new binary merger candidates. We interpret these results in light of eccentric binary formation models. {We rule out formation channels with rates $\gtrsim100$\,Gpc$^{-3}$\,yr$^{-1}$ for $e>0.1$ assuming a black hole mass spectrum with power-law index $\lesssim2$.}
\end{abstract}


\section{Introduction}

In their first two observing runs, the Advanced LIGO and Advanced Virgo detectors discovered 10 binary black hole (BBH) mergers and a binary neutron star merger \citep{2018arXiv181112907T}. These detections have already provided a wealth of information on cosmic processes, including the rate, mass, spin and redshift distribution of BBH mergers \citep{2016ApJ...833L...1A,2018arXiv181112940T}, constraints on general relativity \citep{2016PhRvL.116v1101A,2019arXiv190304467T}, estimates of the Hubble constant \citep{2017Natur.551...85A,2019arXiv190101540T,2019PhRvX...9a1001A}, and constraints on multi-messenger emission from the mergers \citep{2016PhRvD..93l2010A,2016ApJ...826L..13A,2017PhRvD..96b2005A,2018arXiv181002764T,2019arXiv190103310T,2008ApJ...681.1419A}.  

A key question that remains unanswered is how BBHs are formed. Viable formation channels include isolated binary evolution (e.g. \citealt{1998ApJ...506..780B,2002ApJ...572..407B,2014ApJ...789..120B,2016Natur.534..512B,2013ApJ...779...72D,2014A&A...564A.134M,2015MNRAS.451.4086S,2016MNRAS.462.3302E,2016A&A...588A..50M,2016MNRAS.458.2634M,2017MNRAS.472.2422M,2018MNRAS.479.4391M,2017NatCo...814906S,2018MNRAS.480.2011G,2018MNRAS.481.1908K,2018MNRAS.477.4685B}) and dynamical encounters in dense stellar environments, such as globular clusters (e.g. \citealt{2000ApJ...528L..17P,2006ApJ...637..937O,2008ApJ...676.1162S,2010MNRAS.407.1946D,2011MNRAS.416..133D,2015PhRvL.115e1101R,2016PhRvD..93h4029R,2016ApJ...824L...8R,2018ApJ...866L...5R,2017MNRAS.464L..36A,2018PhRvD..97j3014S,2018ApJ...855..124S,2018PhRvL.121p1103F,2019ApJ...871...91Z}), young star clusters (e.g. \citealt{2010MNRAS.402..371B,2014MNRAS.441.3703Z,2016MNRAS.459.3432M,2017MNRAS.467..524B,2018MNRAS.473..909B,2019arXiv190100863D,2018arXiv181106726K}) and galactic nuclei (e.g. \citealt{2009MNRAS.395.2127O,2012ApJ...757...27A,2016ApJ...831..187A,2017ApJ...846..146P,2017MNRAS.464..946S,2017MNRAS.467.4180S,2019arXiv190203242R}). Moreover, the dynamical process known as Kozai--Lidov (KL) resonance \citep{1962AJ.....67..591K,1962P&SS....9..719L} can trigger the merger of a BBH, even if the BBH has not been formed in a dense star cluster. In fact, if the BBH is orbited by a tertiary body (i.e. the BBH is the inner binary of a stable hierarchical triple system), the KL mechanism triggers oscillations of the BBH's eccentricity, which might speed up the merger by gravitational-wave emission. Each channel is expected to produce black hole mergers with different mass and spin distributions \citep{2018arXiv181112940T,2010CQGra..27k4007M,2016ApJ...832L...2R,2016ApJ...818L..22A,2017Natur.548..426F}. The limited statistics from the low number of systems detected through gravitational waves and model uncertainties so far do not allow strong constraints on the formation channels.
 
Orbital eccentricity is a distinguishing feature of dynamical formation channels. Gravitational-wave emission acts to circularize binary orbits by the time they reach orbital frequencies to which Advanced LIGO and Advanced Virgo are sensitive ($\gtrsim10$\,Hz). Eccentric orbits in the Advanced LIGO and Advanced Virgo band indicate either that the binary was formed with small orbital separation and therefore did not have time to circularize, or that some dynamical process increased the eccentricity. For example, KL-induced mergers are expected to be associated with high eccentricities \citep{2017ApJ...841...77A,2019MNRAS.tmp.1958F,2019MNRAS.488...47F,2019MNRAS.486.4781F}. The detection of gravitational waves from an eccentric binary would suggest that binary systems can form dynamically, and could help distinguish between different dynamical formation scenarios (KL oscillations in triple systems or dynamical encounters in dense stellar clusters) \citep{PhysRevD.98.083028}.

In the following we define eccentricity at the time when the gravitational-wave frequency of the binary is at 10\,Hz \citep{1963PhRv..131..435P}. Eccentricity constantly evolves during the inspiral.

Template-based gravitational-wave searches used by Advanced LIGO and Advanced Virgo currently do not include eccentric orbital templates \citep{2018arXiv181112907T}. Quasi-circular waveform templates are able to detect binaries with small eccentricities ($e\lesssim 0.1$), but are inefficient at extracting moderately to highly eccentric binaries \citep{2010PhRvD..81b4007B}. Multiple efforts for generating the full inspiral merger ringdown (IMR) waveforms for the binaries with eccentric orbits are underway \citep{2017PhRvD..96d4028C,2018PhRvD..98d4015H,2019arXiv190403443I,PhysRevD.97.024031,PhysRevD.96.104048}. However, the lack of a reliable and complete waveform model prevents the implementation of a matched-filtering search at this time, and led to the development of alternative search methods \citep{2016PhRvD..93d3007T,2015PhRvD..91f3004C,PhysRevD.98.083028}. 

Here we report the results of a search for eccentric binary black hole mergers with the coherent WaveBurst (cWB) algorithm that does not rely on binary system waveforms. cWB is sensitive to binaries of any eccentricity, and in particular to high-mass black holes. The search has been carried out over data from Advanced LIGO and Advanced Virgo's O1 and O2 observing runs, and found no evidence of eccentric binary signals. This paper evaluates the sensitivity of cWB to eccentric binary mergers, and infers constraints from non-detection on the rate of eccentric mergers.

\section{Detectors and analysis method}
\label{sec:2}
\subsection{Advanced LIGO and Advanced Virgo}
The Advanced LIGO detectors began their first observing run O1 on September 12, 2015, which lasted until January 19, 2016 \citep{2016PhRvX...6d1015A}. During this time they accumulated $T_{\rm obs,1} = 48$\,days of coincident data during which both LIGO Hanford and LIGO Livingston detectors were operating. The second observing run O2 started on November 30, 2016 and lasted until August 25, 2017, resulting in $T_{\rm obs,2} = 118$\,days of coincident data \citep{2018arXiv181112907T}. Advanced Virgo joined the Advanced LIGO detectors on August 1, 2017. The detectors' sensitivity was not uniform during these runs, and there was a marked sensitivity increase from O1 to O2 \citep{2018LRR....21....3A}. As adding Advanced Virgo data was not improving the sensitivity of the search, this analysis only uses data from the Advanced LIGO detectors. 

\subsection{Search description} 

The search for eccentric binary black hole mergers ueses the same configuration of the cWB pipeline \citep{2008CQGra..25k4029K,2016PhRvD..93d2004K} as the binary black hole merger search reported in \cite{2018arXiv181112907T}. An early version of the search is described in \citep{2016PhRvD..93d3007T}. cWB is designed to search for transient signals, without specifying a waveform model. It identifies coherent excess power in multi-resolution time-frequency representations of the detectors' strain data, for signal frequencies up to 1\,kHz and duration up to a few seconds. The excess power is collected in the time-frequency plane assuming monotonically increasing frequency for better collection of the signal energy from binary black holes. The search identifies events that are coherent in multiple detectors and reconstructs the source sky location and signal waveforms by using the constrained maximum likelihood method. 

The cWB detection statistic $\rho$ is based on the coherent energy $E_{\rm c}$ obtained by cross-correlating the signal waveforms reconstructed in the network of detectors. It is proportional to the coherent network signal-to-noise ratio. %This detection statistic also accounts for the quality of the chirp reconstructed by fitting the time-frequency pixels with the Newtonian chirp mass formula, the quality of the chirp is defined in \citep{chirp_V}. The chirp mass $\mathcal{M}$ of a binary with black holes of masses $m_1$ and $m_2$ is defined as $\mathcal{M}\equiv (m_{1}m_{2})^{3/5}(m_1+m_2)^{-1/5}$.  
The estimation of statistical significance of an event is done by ranking the $\rho$ of the event against the $\rho$ distribution for background events obtained by repeating the analysis on time-shifted
data. To exclude astrophysical events from the background sample, the time
shifts are much larger than the expected signal delay
between the detectors. Each cWB event was assigned a False Alarm Rate (FAR) based on the rate of background triggers with $\rho$ higher than that of the event.  

To increase the robustness against non stationary detector noise generating glitches, cWB uses signal-independent vetoes: the network correlation $c_{\rm c}=E_{\rm c}/(E_{\rm c}+E_{\rm n})$, where $E_{\rm n}$ is the residual noise energy estimated after the reconstructed signal pixels are subtracted from the data. For a gravitational-wave signal we expected $c_{\rm c}\approx1$ while for glitches $c_{\rm c} \ll1$. Events with $c_{\rm c} < 0.7$ are rejected.

Detector characterization studies are also carried out to ensure that candidate events are not due to instrumental or environmental artifacts. We have rejected the times where significant instrumental artifacts make the data unusable \citep{2016CQGra..33m4001A}.

\subsection{Simulated astrophysical signals} 
\label{sec:simsig}

\begin{figure*}
  \centering
   \includegraphics[width=\textwidth]{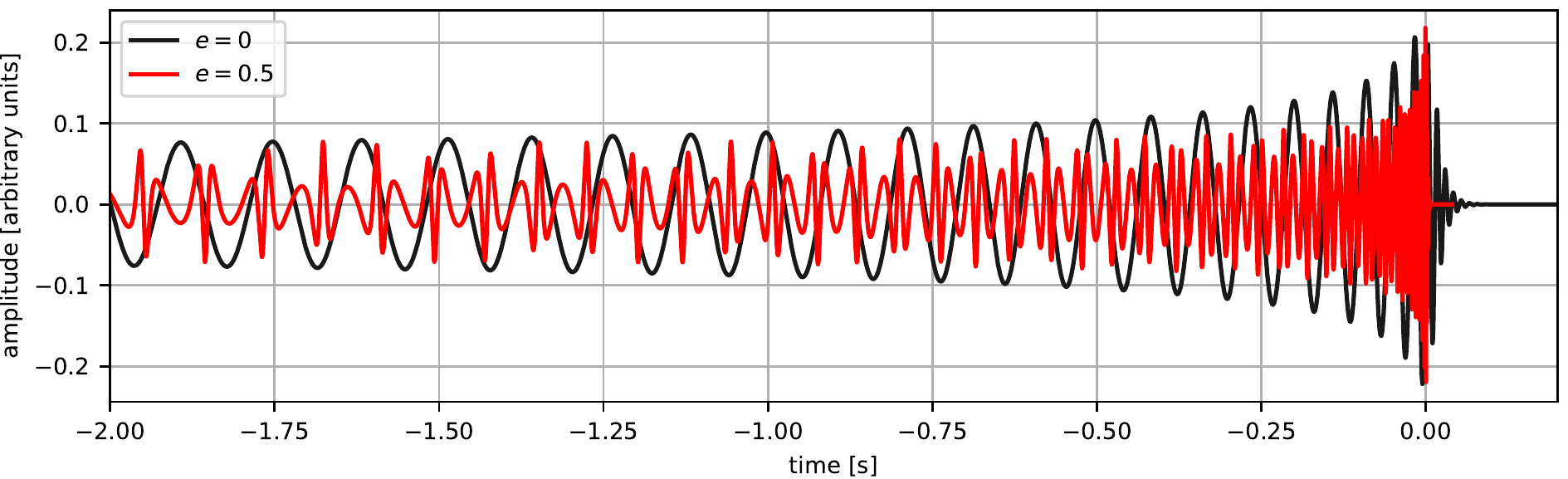}
   \caption{Examples of gravitational waveforms for a 10\,M$_\odot - 10$\,M$_{\odot}$ BBH system with eccentricities 0 (black) and 0.5 (red).} 
   \label{fig:waveform}
\end{figure*}

In order to estimate the sensitivity of our search, we simulated eccentric BBH signals, injected them into detector data and searched for them using cWB. We used a BH mass range of 5\,M$_\odot - 50\,$M$_\odot$ \citep{2018arXiv181112940T}, and eccentricities in the $e\in[0,0.99]$ range. We assumed that BHs have zero spin. These simulations were carried out to quantify the search sensitivity for individual binaries. Below we considered different mass distributions to characterize our sensitivity. 

At the time of the analysis only one set of templates was available for the generation of full inspiral-merger-ringdown eccentric binary waveforms including generic spin configurations by \cite{emlp2013}.
It uses a prescription based on the equations of motion of a geodesic in a Kerr spacetime, coupled with the quadrupole formula for the gravitational radiation. The model defines an effective Kerr spacetime whose mass and spin parameters are set equal to the total mass and orbital angular momentum of the binary. The binary is evolved based on the behavior of a timelike geodesic in the effective Kerr spacetime, but the mass and angular momentum of this spacetime are changed at each time step based on the emitted energy and angular momentum calculated in the quadrupole approximation. This approach reproduces the correct orbital dynamics in the Newtonian limit and general-relativistic test particle limit. This model also incorporates strong-field features such as pericenter precession, frame dragging, and the existence of unstable orbits and related zoom-whirl dynamics \citep{emlp2013}. The inspiral waveforms obtained using the above treatment are stitched to a merger model that was developed for quasicircular mergers but also performs well for eccentric mergers with little modification \citep{Baker:2008mj,Kelly:2011bp}. In Fig. \ref{fig:waveform} we show this waveform for the case of circular and eccentric ($e=0.5$) orbits.

{The waveforms we used here to simulate gravitational waves from eccentric binaries are approximate. Compared to gravitational waveforms obtained using general-relativistic numerical simulations, the waveforms differ in overlap by up to a few percent \citep{emlp2013}. However, we found that cWB can detect the waveforms used here with equal sensitivity to precise waveforms obtained using numerical simulations, making their use appropriate to characterize search sensitivity. This is due to the fact that cWB does not rely on the precise gravitational waveform, making a few percent difference negligible.}

%The waveforms we used here are likely not sufficiently accurate for optimal template-based detection. Compared to gravitational waveforms obtained using general-relativistic numerical simulations, the waveforms differ in overlap by up to a few percent \citep{emlp2013}. However, these waveforms are sufficiently accurate to be used to assess the efficiency of the unmodeled search used. %The difference between the estimated effective orbital radii with this waveform model at zero eccentricity and an accurate IMR model EOBNRv2 is no more than 5\% in the parameter space we have considered.  

\section{Results} 
\label{sec:results}

This search has detected 7 of the 10 BBH events that were identified by template based searches (its sensitivity compared to template based searches is higher for higher mass binaries; see Table 1 in \citealt{2018arXiv181112907T}). We considered these events to have no eccentricity. Our search did not detect any gravitational-wave event beyond these. Therefore, we concluded that no eccentric BBH merger has been detected. Below we present our search sensitivity to interpret this non-detection. 

{We note that the detection of only 7 out of 10 BBH events by cWB is consistent with its relative sensitivity compared to template-based searches, which are more sensitive in particular at low black hole masses (see Fig. \ref{fig:sensitivedistance}). Our interpretation below does not depend on, or make use, of this fractional overlap between the two search types.}

We further note that the detection by cWB and the less-confident template-based detection of an event would not necessarily mean that the event was an eccentric binary (see, e.g., \citealt{2018arXiv181112907T}). The eccentricity of a detected event would need to be independently measured (e.g., \citealt{PhysRevD.98.083028}).

\subsection{Sensitivity to eccentric mergers}

\begin{figure}[htb!]
  \centering
   \includegraphics[width=0.49\textwidth]{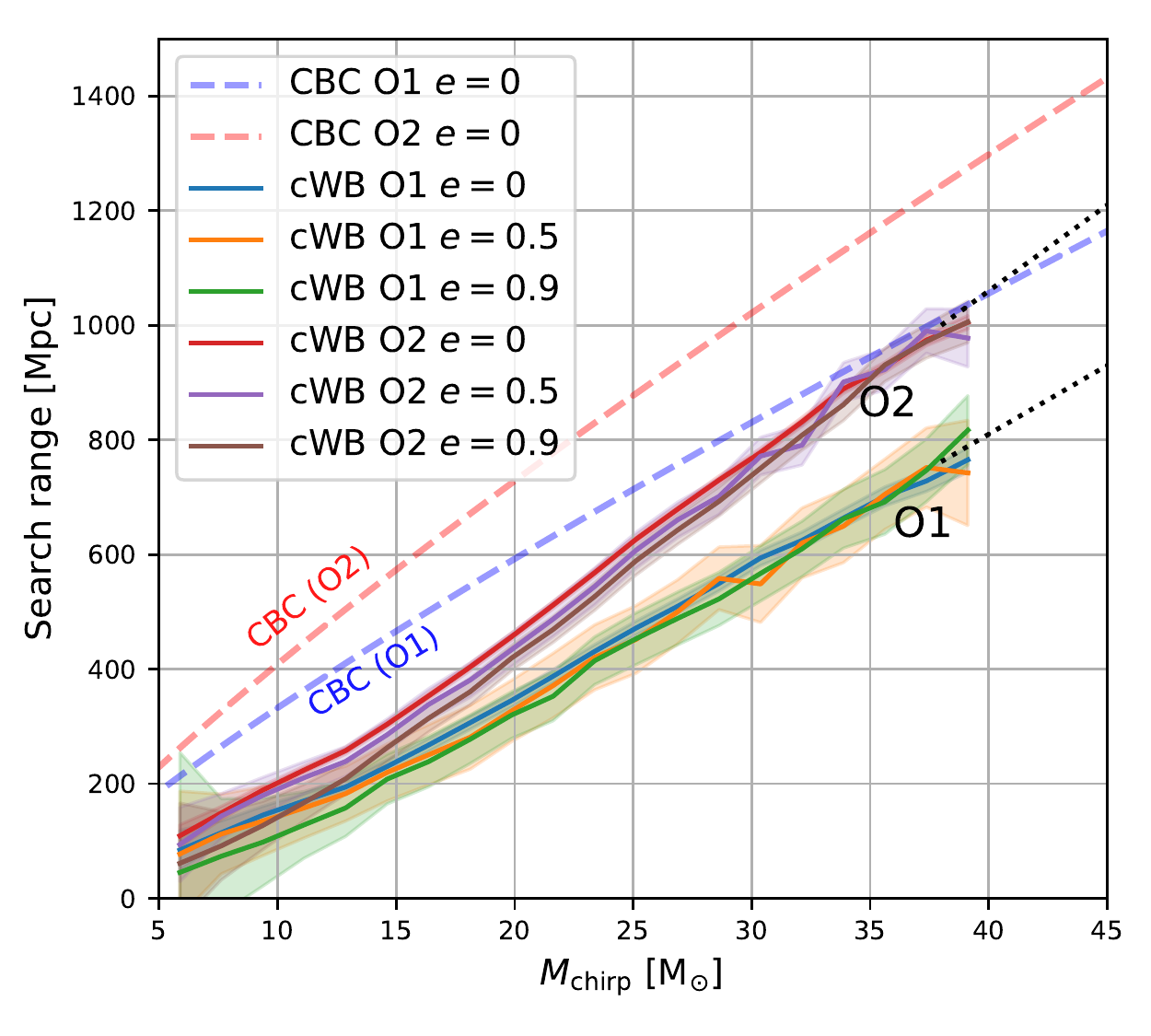}
   \caption{Range of the cWB analysis to BBH mergers as a function of the binary's chirp mass, separately for the O1 and O2 observing runs, and for different orbital eccentricities (see legend). The shaded regions represent $1\sigma$ uncertainties. The dotted lines are linear fits on the ranges at chirp masses $>30$\,M$_{\odot}$ for $e=0$. For comparison, we show the sensitive ranges for template-based searches for compact binary coalescence (CBC), assuming $e=0$, for O1 and O2 \citep{2018LRR....21....3A}. Masses are given in source-frame.} 
   \label{fig:sensitivedistance}
\end{figure}

We characterized the sensitivity of our search by calculating its {\it range}---the distance, averaged over observation time, sky location and orientation, within which a BBH can be detected with false alarm rate $\leq10^{-2}$\,yr$^{-1}$. For this calculation we adopted a standard cosmological model with Hubble parameter $H_0 = 67.9$\,km\,s$^{-1}$Mpc$^{-1}$ and $\Omega_{\rm M} = 0.3065$ \citep{2016A&A...594A..13P}. The range depends on the black hole masses, and is different for the O1 and O2 observing runs. In particular it depends on the chirp mass $\mathcal{M}$ of the binary, where $\mathcal{M}\equiv (m_{1}m_{2})^{3/5}(m_1+m_2)^{-1/5}$ for black hole masses $m_1$ and $m_2$. We find that cWB range is independent of the eccentricity for the whole mass range considered (see also \citealt{2016PhRvD..93d3007T}). Our ranges, using the eccentric waveforms described in Section \ref{sec:simsig}, are shown in Fig. \ref{fig:sensitivedistance}. We additionally see that the sensitive range of cWB grows faster with chirp mass than the range of template-based searches, making cWB additionally useful for circular binaries at higher masses (see also \citealt{2017PhRvD..96b2001A}).

\subsection{Astrophysical constraints}

In order to compare our results to astrophysical source populations, we calculated the volume-time (VT) probed by our search. VT depends on the mass distribution of the BBH population. Dynamical formation channels are expected to result in different BH mass and mass ratio distributions than BBH mergers from field binaries \citep{2018PhRvD..98l3005R,2016MNRAS.463.2443K}. We considered a BBH mass distribution such that the mass of the more massive BH, $m_{1}$, follows a power-law distribution $m^{-\beta}$ within the range $[5\,$M$_\odot,50$M$_\odot]$ for different $\beta$ values (see below), while the second BH's mass, $m_{2}$ is uniformly distributed within the range $[5\,$M$_\odot,m_1]$. The mass distribution of BBH mergers detected by Advanced LIGO and Advanced Virgo so far is somewhat different from this assumed distribution \citep{2018arXiv181112907T,2018arXiv181112940T}. However, eccentric BBH merger channels are likely responsible for only a subset of these observations and therefore they do not fully determine the overall spectrum. With this mass distribution model, we find that $\mbox{VT}(\beta)\approx\{6.6,2.4,0.75\}\times 10^{-2}$\,Gpc$^{3}$\,yr for $\beta=\{1,2,3\}$, respectively. %The corresponding rate density upper limits at 90\% confidence level are $\{30,90,300\}$\,Gpc$^{-3}$\,yr$^{-1}$, respectively. These upper limit values are rounded to the first significant digit to reflect that they are approximations due to modeling uncertainties. We find that this result does not depend on the eccentricity distribution of the source population as our search sensitivity only weakly depends on eccentricity.

The BBH merger rate density for processes that can lead to eccentric orbits in the Advanced LIGO and Advanced Virgo band is mostly predicted to be up to a few \,Gpc$^{-3}$\,yr$^{-1}$ \citep{2016ApJ...831..187A,2016ApJ...824L...8R,2017ApJ...836...39S,2017ApJ...835..165B,2017ApJ...846..146P,2018ApJ...856..140H,2018ApJ...865....2H,2019arXiv190301405Y,2006ApJ...648..411K}, while some more extreme models predict merger rate densities up to 100\,Gpc$^{-3}$\,yr$^{-1}$ \citep{2016ApJ...828...77V,2019arXiv190203242R}. The fraction of mergers from these processes that have high eccentricity ($e\gtrsim0.1$) ranges from $\sim1\%$ \citep{2018ApJ...853...93R,2018ApJ...864..134R} to close to all mergers \citep{2017ApJ...846..146P,2018ApJ...860....5G}. 

In order to understand the astrophysical rate density constraints of our results, we considered a dynamical formation channel that produces BBH mergers at rate density $R_{\rm dyn}$, with a mass power-law index $\beta$ (see above). We assumed that a fraction $f_{\rm ecc}$ of mergers from this channel have eccentricities $e>0.1$, and that this BBH sub-population follows the mass distribution considered here. We further assumed that all BBH mergers detected by Advanced LIGO and Advanced Virgo so far have eccentricities $e<0.1$. The expected number of eccentric mergers ($e>0.1$) from this model detected by cWB during O1 and O2 is then 
\begin{equation}
\langle N_{\rm cWB,ecc}\rangle = R_{\rm dyn} f_{\rm ecc} \mbox{VT}(\beta).
\end{equation}
Given that no such eccentric merger was detected, the Neyman 90\% confidence-level upper limit is
\begin{equation}
R_{\rm ul,ecc} = \frac{2.3}{f_{\rm ecc} \mbox{VT}(\beta)}.
\end{equation}
where $\beta=\{1,2,3\}$. We obtained 
\begin{equation}
R_{\rm ul,ecc}=\{30,90,300\} \, f_{\rm ecc}^{-1}\,\,\mbox{Gpc}^{-3}\,\mbox{yr}^{-1}
\end{equation}
for $\beta~=~\{1,2,3\}$, respectively. The quoted approximate values were rounded to the first significant digit. We found that this result does not depend on the eccentricity distribution of the source population as our search sensitivity only weakly depends on eccentricity.

Our results rule out models predicting $\gtrsim100$\,Gpc$^{-3}$\,yr$^{-1}$ merger rate densities \citep{2016ApJ...828...77V,2019arXiv190203242R} for $\beta\lesssim2$ if the majority of mergers in the given model have eccentricities $e>0.1$, {while the results are consistent with a number of other models (e.g., \citealt{2017ApJ...841...77A,2019MNRAS.488.2825F,Antonini:2012ad}).}

\section{Conclusion}
\label{sec:conclusion}

We searched for eccentric BBH mergers using the cWB algorithm. We showed that the sensitivity of our method is independent of the eccentricity at the time the binary enters Advanced LIGO and Advanced Virgo's frequency band at $\sim10$\,Hz. 

Our search only uncovered binaries that have also been found by template-based searches that do not appear to have eccentric orbits. We interpreted this non-detection in light of the expected merger rate density of BBH formation channels that can produce eccentric orbits, and the fraction of these mergers that have eccentricities $\gtrsim0.1$. Our results rule out the highest end of the rate density predictions ($\gtrsim100$\,Gpc$^{-3}$\,yr$^{-1}$) assuming that the majority of the binaries from these channels have $e>0.1$, and that the power-law index of the BH mass spectrum is $\lesssim2$.

Future observing runs by Advanced LIGO, Advanced Virgo and KAGRA \citep{PhysRevD.88.043007} will provide substantially improved sensitivity to probe formation mechanisms resulting in eccentric binaries \citep{2018LRR....21....3A}. %Additionally, advances in theoretical predictions on the expected parameters of dynamical binary formation will also be critical in connecting observations with astrophysical processes.

The authors gratefully acknowledge the support of the United States
National Science Foundation (NSF) for the construction and operation of the
LIGO Laboratory and Advanced LIGO as well as the Science and Technology Facilities Council (STFC) of the
United Kingdom, the Max-Planck-Society (MPS), and the State of
Niedersachsen/Germany for support of the construction of Advanced LIGO 
and construction and operation of the GEO600 detector. 
Additional support for Advanced LIGO was provided by the Australian Research Council.
The authors gratefully acknowledge the Italian Istituto Nazionale di Fisica Nucleare (INFN),  
the French Centre National de la Recherche Scientifique (CNRS) and
the Foundation for Fundamental Research on Matter supported by the Netherlands Organisation for Scientific Research, 
for the construction and operation of the Virgo detector
and the creation and support  of the EGO consortium. 
The authors also gratefully acknowledge research support from these agencies as well as by 
the Council of Scientific and Industrial Research of India, 
the Department of Science and Technology, India,
the Science \& Engineering Research Board (SERB), India,
the Ministry of Human Resource Development, India,
the Spanish  Agencia Estatal de Investigaci\'on,
the Vicepresid\`encia i Conselleria d'Innovaci\'o, Recerca i Turisme and the Conselleria d'Educaci\'o i Universitat del Govern de les Illes Balears,
the Conselleria d'Educaci\'o, Investigaci\'o, Cultura i Esport de la Generalitat Valenciana,
the National Science Centre of Poland,
the Swiss National Science Foundation (SNSF),
the Russian Foundation for Basic Research, 
the Russian Science Foundation,
the European Commission,
the European Regional Development Funds (ERDF),
the Royal Society, 
the Scottish Funding Council, 
the Scottish Universities Physics Alliance, 
the Hungarian Scientific Research Fund (OTKA),
the Lyon Institute of Origins (LIO),
the Paris \^{I}le-de-France Region, 
the National Research, Development and Innovation Office Hungary (NKFIH), 
the National Research Foundation of Korea,
Industry Canada and the Province of Ontario through the Ministry of Economic Development and Innovation, 
the Natural Science and Engineering Research Council Canada,
the Canadian Institute for Advanced Research,
the Brazilian Ministry of Science, Technology, Innovations, and Communications,
the International Center for Theoretical Physics South American Institute for Fundamental Research (ICTP-SAIFR), 
the Research Grants Council of Hong Kong,
the National Natural Science Foundation of China (NSFC),
the Leverhulme Trust, 
the Research Corporation, 
the Ministry of Science and Technology (MOST), Taiwan
and
the Kavli Foundation.
The authors gratefully acknowledge the support of the NSF, STFC, MPS, INFN, CNRS and the
State of Niedersachsen/Germany for provision of computational resources.

\bibliography{main}

\end{document}